\documentclass[11pt,twoside]{article}
\usepackage{asp2006}
\usepackage{psfig}
\usepackage{epsf}
\usepackage{graphics}
\usepackage{lscape}
\begin{document}
\title{Correlation Analysis of Mode Frequencies with Activity Proxies at Different
Phases of the Solar Cycle}
\author{Kiran Jain,
S. C. Tripathy,
and F. Hill}
\affil{National Solar Observatory, 950 N. Cherry Avenue, Tucson, AZ  85719, USA. Email: kjain@noao.edu}

\begin{abstract}
We analyze intermediate degree {\it p}- and {\it f}-mode eigenfrequencies 
measured by GONG and MDI/SOHO for a complete solar cycle to study
their correlation with solar activity. We demonstrate that the frequencies do 
vary linearly with the activity, however the degree of correlation differs 
from phase to phase of the cycle. During the rising and the declining phases, 
the mode frequencies are strongly correlated with the activity proxies whereas 
during the low- and high-activity periods, the frequencies have significantly 
lower correlation with all the activity proxies considered here. 
\end{abstract}

\section{Introduction}
The linear variation of mode frequencies with the changing solar magnetic  
activity is well established \citep{jb03,wjc07}. However, detailed studies based on 
high quality uniform data indicate the complexity of the relationship between 
mode frequencies and solar activity. For example, \citet{howe99} have
shown that there is a latitudinal variation in frequencies and splitting coefficients. 
In addition, \citet{sct07} found a  year-wise distribution in linear-regression slopes 
(i e. the change in frequency per unit change in activity) and the degree of correlation.
Thus, with the availability of continuous data for a complete solar cycle, it is 
important to study correlations of oscillation frequencies with different measures 
of the solar activity in order to understand the source of the variability.

\section{Analysis and Results}
The analysis presented here uses oscillation data sets obtained from the Global 
Oscillation Network Group (GONG) and Michelson Doppler Imager (MDI) {\it onboard
Solar and Helipspheric Observatory (SOHO)} and solar activity data. It covers 
a period of about 13 years, i.e. a complete solar cycle 
including both minimum phases at the beginning and end of solar cycle 23. The 128 108-day
GONG data sets spanning over the period from 1995 May 7 to 2008 Feb 27 are continuous
while the 61 72-day MDI 
data sets have two large gaps in 1998-1999. The MDI data cover the period from 
1996 May 1 to 2008 Sept 30. The activity data used here are: the integrated emission
from the solar disk at 10.7 cm wavelength ($F_{10}$), the line-of sight magnetic field 
strength from Kitt Peak Observatory (KPMI), the international sunspot number ($R_I$),
the Mt. Wilson sunspot index (MWSI), the magnetic plage strength index (MPSI) from Mt. 
Wilson and the total solar irradiance (TSI).  Details of the various data sets are given
in \citep{jain08}.

The number of {\it p-} modes analysed here are 479 for GONG and 876 for MDI data sets in the frequency range 1.5 $\le \nu \le$ 4.0 mHz. These modes are observed in all data sets of GONG or MDI. We also analyze 76 {\it f}-modes observed in all MDI data sets. 
 The frequency shifts ($\delta\nu$) are calculated with respect to the reference frequency which is determined by taking an average of the frequencies of a particular multiplet ($n,\ell$).

\begin{figure}[!t]
\plotone{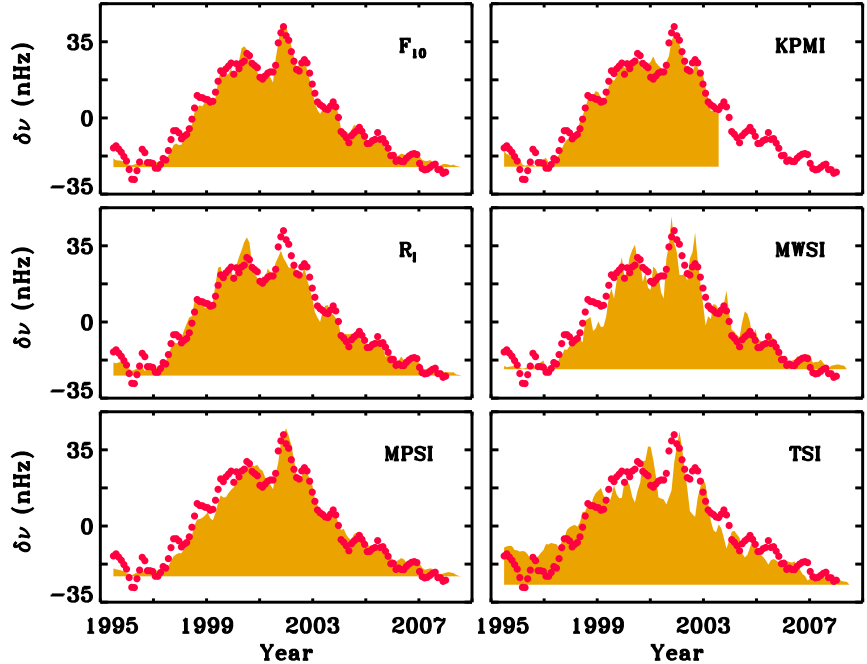}
\caption{\label{fig:fig1}Temporal evolution of the GONG {\it p}-mode frequency shifts (symbols) with different activity proxies (filled regions).}
\end{figure}

The temporal variation of GONG frequency shifts ($\delta\nu$) with various measures of solar activity ($I$) are shown in Figure~1. It is evident that the frequency shifts follow the general trend of the solar activity. The correlation coefficients between $\delta\nu$ and $I$ obtained in all cases are comparable. However, we find significant different correlation coefficients when we divide the activity cycle into four phases as shown in Figure 2; the periods of minimum activity at the beginning and end of the solar cycle ({\it Phase I}), rising activity ({\it Phase II}), high-acivity ({\it Phase III}), and declining activity ({\it Phase IV}). In the right panel of Figure~2, we compare the phase-wise Pearson's linear correlation coefficients ($r_P$) for both GONG and MDI data sets with those obtained for the complete cycle.  It is interesting to note that the correlation between $\delta\nu$ and solar activity changes significantly from phase to phase; the rising and declining phases are better correlated than the low-
and high-activity phases. The frequencies during {\it Phase I} do not correlate well with any of the proxies.  During {\it Phase III}, we obtain significant correlations for $F_{10}$ and KPMI while a substantial decrease is noticed for $R_I$, MWSI and TSI. 

Figure~3 (left panel) shows the temporal variation of {\it f}-mode frequencies. We notice two distinct features in frequency shifts; the persisting strong 1-year periodicity as discussed by several authors \citep{antia01,jb03,dzi05} and the frequencies at current solar minimum (2007-2008) are lower than those at the previous minimum (1996). The correlation coefficients obtained with the original and smoothed frequency shifts are also shown in the right panel. The smoothed frequency shifts are the running mean of five points to minimize the effect of 1-year periodicity. It is seen that smoothing enhances the correlation in all cases. The variation in correlation coefficients for {\it f}-modes at different phases are consistent with those for {\it p}-modes. In both cases, MDI data sets have good correlation with TSI at the low-activity phase that requires a detailed investigation. 
 
\begin{figure}[!t]
\plottwo{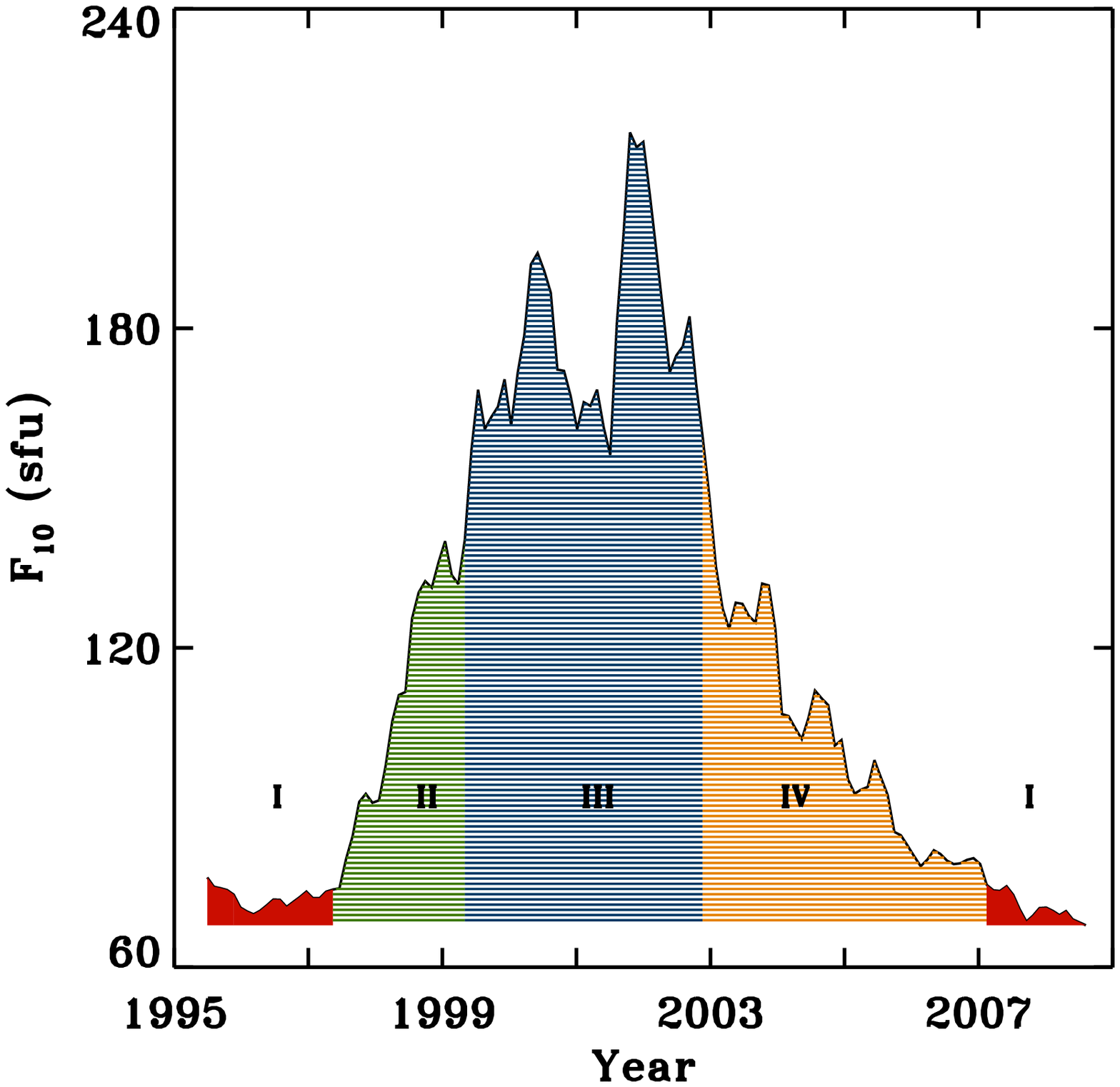}{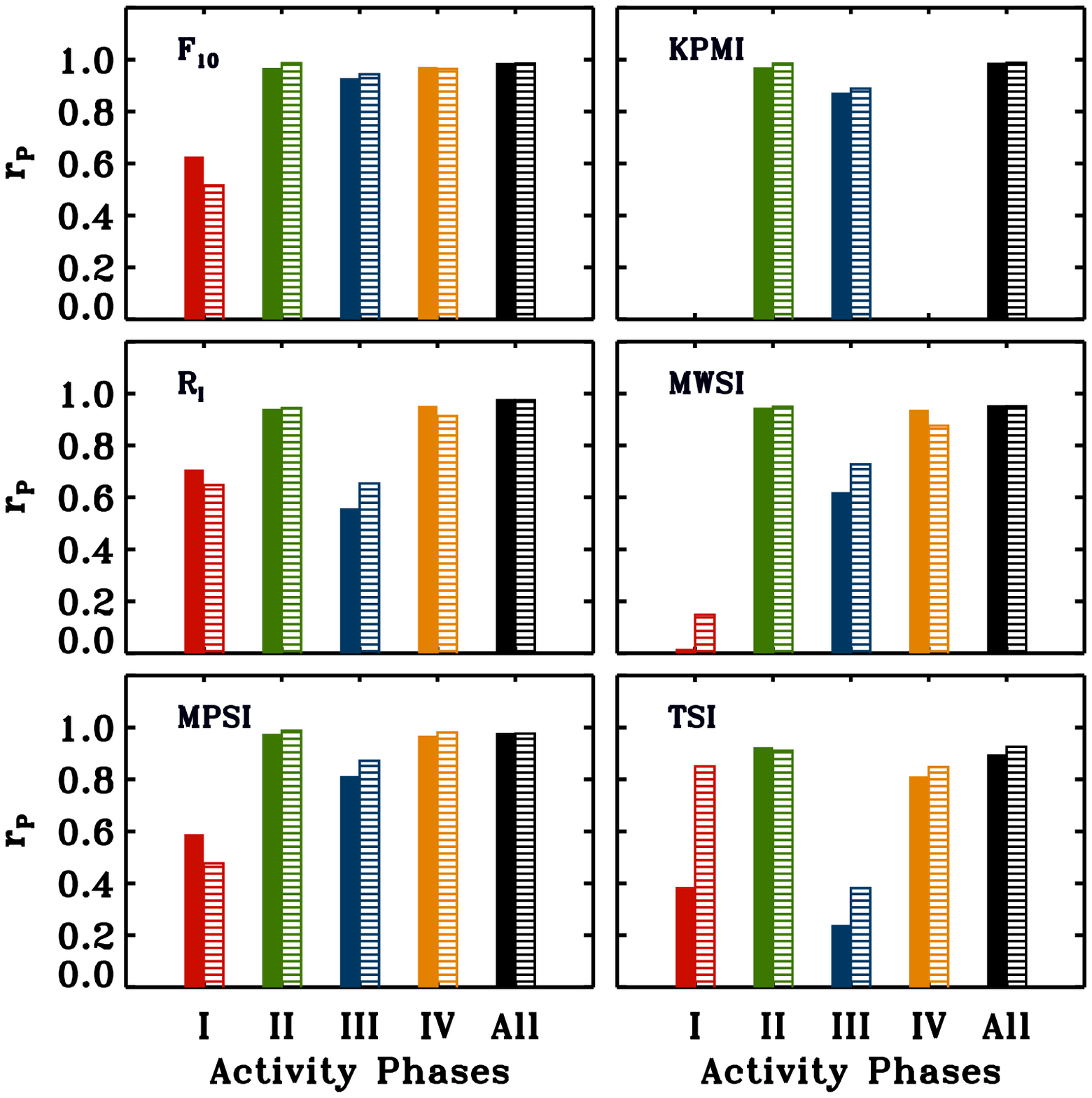}
\caption{\label{fig:fig2}(Left) Different phases of the solar activity cycle: Phase I shown by filled regions at the begining and the end of curve corresponds to low activity period, Phase II the rising activity, Phase III the high activity and Phase IV the declining activity periods. (Right) Bar-chart showing phase-wise variation of the Pearson's linear correlation coefficient between {\it p}-mode frequency shifts and different activity indices. Each activity phase has values for GONG (filled) and MDI (hatched) data sets. The missing values for KPMI are due to unavailablity of sufficient activity data points. Correlation coefficients for all data sets are also shown here. } 
\end{figure}

\section{Summary}
In summary, the improved and continuous measurements of intermediate-degree mode frequencies for a complete solar cycle demonstrate that, while the frequencies vary in phase with the solar activity, the degree of correlation between frequencies and activity indices differs from one activity phase to another. Although there is a strong correlation during rising and declining activity phases, we find a significant decrease in correlation at the low- and high-activity phases. 

\begin{figure}[!t]
\plottwo{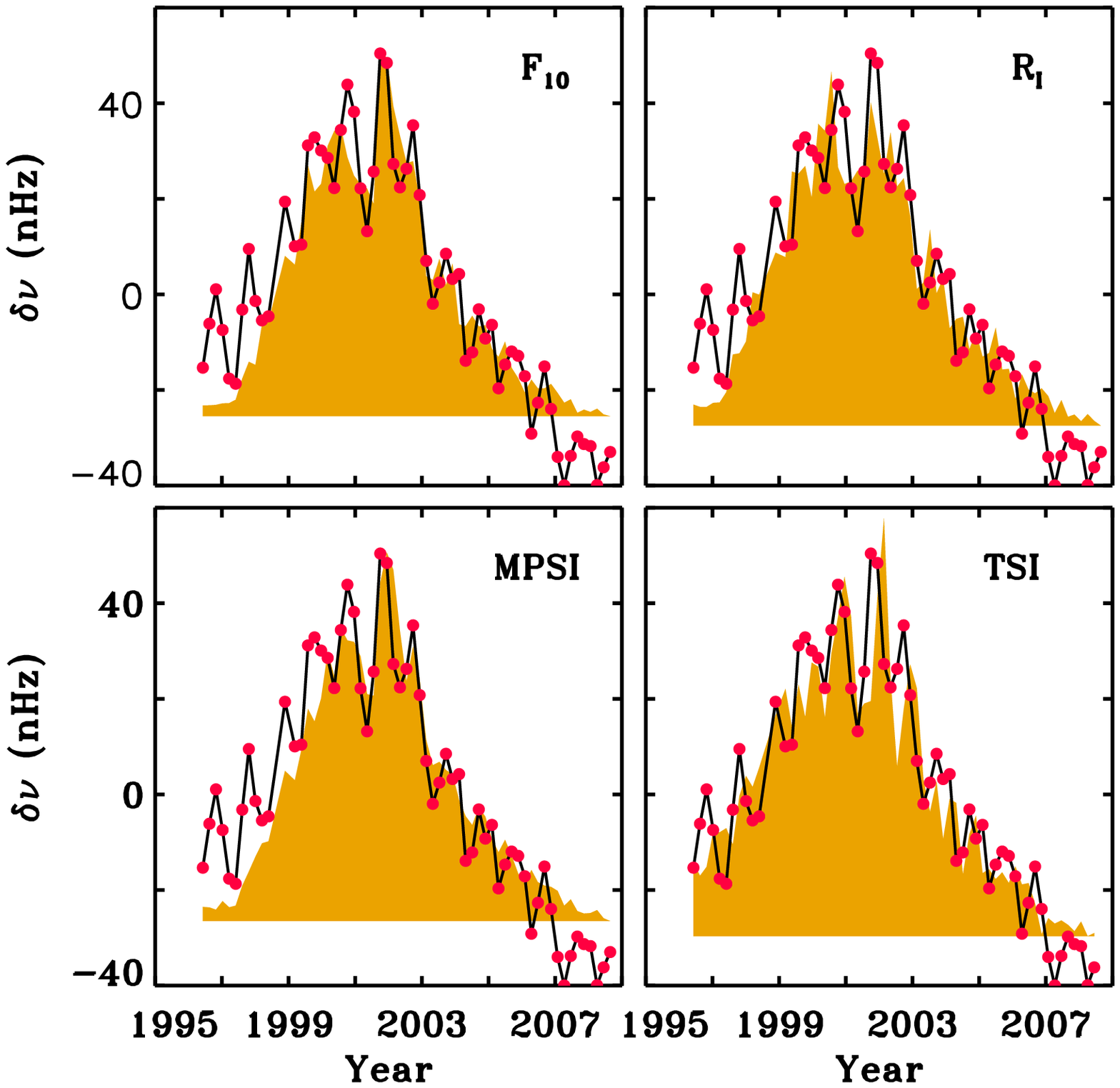}{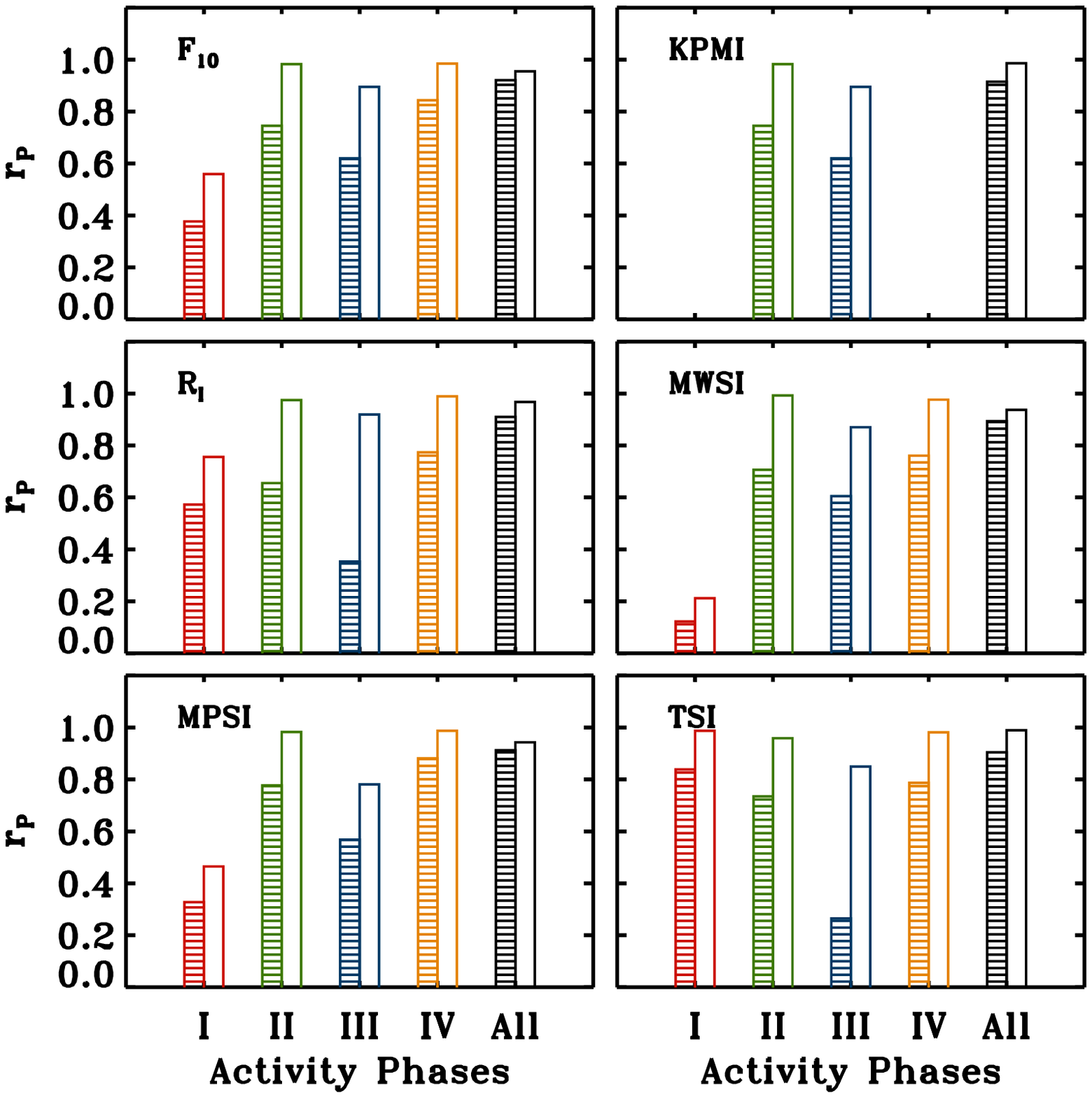}
\caption{\label{fig:fig3} (Left) Temporal evolution of {\it f}-mode frequency shifts (circles) and solar activity (filled regions). (Right) Linear correlation coefficients between {\it f}-mode frequency shifts and activity indices for different phases of the solar cycle. Hatched bars show the correlation coefficients between actual values of frequency shifts and activity proxies while open bars are for correlation coefficients between smoothed value of frequency shifts and activity proxies.}

\end{figure}

\acknowledgements{
This work utilizes data obtained by the Global Oscillation Network
Group (GONG) project, managed by the National Solar Observatory, which
is operated by AURA, Inc. under a cooperative agreement with the
National Science Foundation. The data were acquired by instruments
operated by the Big Bear Solar Observatory, High Altitude Observatory,
Learmonth Solar Observatory, Udaipur Solar Observatory, Instituto de
Astrof\'{\i}sico de Canarias, and Cerro Tololo Interamerican
Observatory. It also utilises data from the Solar Oscillations 
Investigation/Michelson Doppler Imager on the Solar and Heliospheric 
Observatory. SOHO is a mission of international cooperation
 between ESA and NASA. NSO/Kitt Peak magnetic used here are 
produced cooperatively by NSF/NOAO; NASA/GSFC and NOAA/SEL. 
 This study also includes data from the synoptic program at 
the 150-Foot Solar Tower of the Mt. Wilson Observatory, operated by 
UCLA, with funding from NASA, ONR and NSF, under agreement with the 
Mt. Wilson Institute. The unpublished solar irradiance dataset 
(version v6\_001\_0804) was obtained from VIRGO Team through 
PMOD/WRC, Davos, Switzerland. This work
was supported by NASA grants NNG05HL41I and NNG08EI54I.}

\end{document}